\documentclass[twocolumn,superscriptaddress,amsmath,amssymb,prl,longbibliography]{revtex4-1}

\ifx\pdfoutput\@undefined\usepackage[usenames,dvips]{color}
\else\usepackage[usenames,dvipsnames]{color}
\usepackage[pdftex]{graphicx}
\usepackage{bm}
\usepackage{amssymb}
\usepackage{hyperref}
\usepackage{color}

\usepackage{amsmath}
\usepackage{eufrak}

\usepackage[T1]{fontenc}
\usepackage[latin1]{inputenc}

\renewcommand{\vec}[1]{\mathbf{#1}}

\newcommand{\eps}{\varepsilon}

\newcommand{\CKB}{\textcolor{black}}

\usepackage{graphicx}
\usepackage{color}

\begin{document}

\title{Klein-Gordon Representation of Acoustic Waves \\ and
Topological Origin of Surface Acoustic Modes}

\author{Konstantin Y. Bliokh}
\affiliation{Theoretical Quantum Physics Laboratory, RIKEN Cluster for Pioneering Research, Wako-shi, Saitama 351-0198, Japan}
\affiliation{Nonlinear Physics Centre, RSPE, The Australian National University, Canberra, ACT 0200, Australia}

\author{Franco Nori}
\affiliation{Theoretical Quantum Physics Laboratory, RIKEN Cluster for Pioneering Research, Wako-shi, Saitama 351-0198, Japan}
\affiliation{Physics Department, University of Michigan, Ann Arbor, Michigan 48109-1040, USA}


\begin{abstract}
Recently, it was shown that surface electromagnetic waves at interfaces between continuous homogeneous media (e.g., surface plasmon-polaritons at metal-dielectric interfaces) have a topological origin [K. Y. Bliokh {\it et al}., Nat. Commun. {\bf 10}, 580 (2019)]. This is explained by the nontrivial topology of the non-Hermitian photon {\it helicity} operator in the {\it Weyl-like} representation of Maxwell equations. Here we analyze another type of classical waves: longitudinal acoustic waves corresponding to spinless phonons. We show that surface acoustic waves, which appear at interfaces between media with opposite-sign densities, can be explained by similar topological features and the bulk-boundary correspondence. However, in contrast to photons, the topological properties of sound waves originate from the non-Hermitian {\it four-momentum} operator in the {\it Klein-Gordon} representation of acoustic fields.
\end{abstract}

\maketitle


{\it Introduction.---}
Maxwell electromagnetism, elasticity, and acoustics describe various classical waves via different types of wave equations \cite{Jackson,Auld,LLFluid}. Analogies between these waves are very fruitful and repeatedly resulted in the mutual export of ideas between optics and acoustics. To name a few, acoustic crystals/metamaterials \cite{Lu2009,Hussein2014,Ma2016,Cummer2016}, vortex beams \cite{Volke2008,Demore2012,Anhauser2012}, and topological systems \cite{Yang2015,Wang2015,Nash2015,He2016,Ni2019} were developed in parallel with their optical counterparts \cite{Joannopoulos1997,Smith2004,Alu2007,Bliokh2008,Allen1999,Molina2007,
Franke2008,Bliokh2015,Wang2009,Rechtsman2013,Khanikaev2013,Lu2014,
Ozawa2018} and attracted great attention in the past decades.

Surface waves at interfaces between continuous media, such as surface plasmon-polaritons, are highly important for modern optics \cite{Smith2004,Alu2007,Bliokh2008,Zayats2005,Maier2007,Shadrivov2004,Kats2007}. However, there are only few works analyzing acoustic analogues of such waves \cite{Ambati2007,Christensen2007,Park2011,Bliokh2019II}. The reason for this is that such waves (for linear longitudinal sound fields) appear only at interfaces with {\it negative-density} media, i.e., acoustic {\it metamaterials} \cite{Ma2016,Cummer2016}. Surface electromagnetic waves also require media with negative parameters (permittivity or permeability), but there are natural media with such parameters: e.g., metals. Nonetheless, here we explore the fundamental origin of surface acoustic modes, and show that this reveals nontrivial intrinsic properties of the acoustic wave equations.

Surface modes are particularly important in the context of topological quantum or classical-wave systems, which are currently attracting enormous attention 
\cite{Lu2014,Ozawa2018,Hasan2010,Qi2010,Qi2011,Chiu2016}. Indeed, the topological approach provides a direct link between the nontrivial intrinsic properties of the bulk Hamiltonian and the appearance of surface modes at interfaces between topologically-different media.

\begin{figure}
\includegraphics[width=\columnwidth]{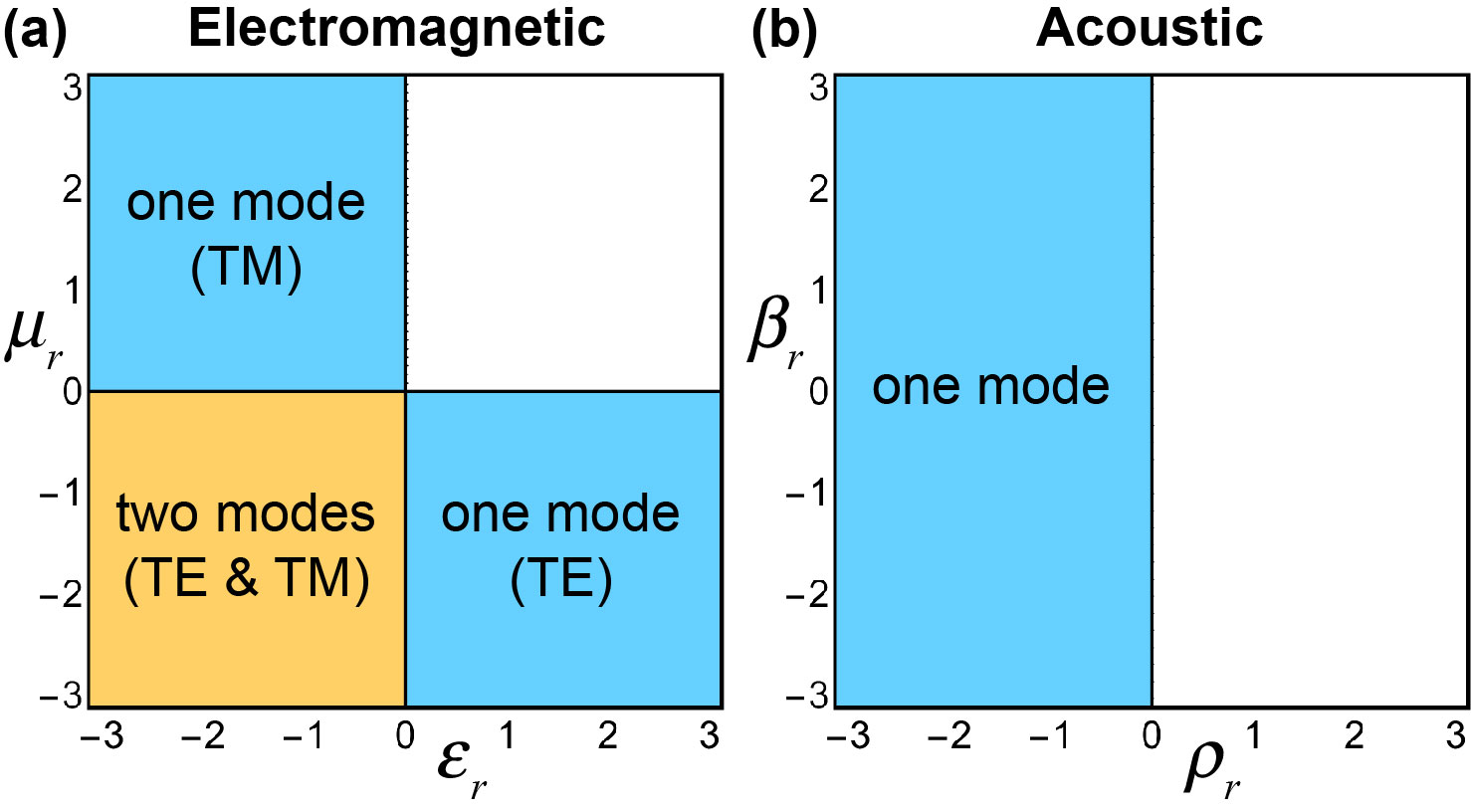}
\caption{Phase diagrams showing the domains of existence of the electromagnetic (a) and acoustic (b) surface modes at interfaces between continuous media. The electromagnetic diagram is explained in \cite{Bliokh2019} ($\eps_r$ and $\mu_r$ are the relative permittivity and permeability of the two media), while the acoustic one is described by the topological $\mathbb{Z}_2$ index (10) and bulk-boundary correspondence (11) ($\rho_r$ and $\beta_r$ are the relative density and compressibility of the two media). 
\label{Fig1}}
\end{figure}

Recently we have shown \cite{Bliokh2019} that electromagnetic surface waves and their domains of existence can be explained by the fundamental topological origin and the bulk-boundary correspondence. Namely, analyzing the relativistic Weyl-like form of Maxwell equations, one can see that the photon {\it helicity} operator: (i) is generally non-Hermitian, \CKB{even in idealized lossless media,} and (ii) has a nontrivial topological structure involving two medium parameters: the permittivity $\eps$ and permeability $\mu$. According to this approach, the helicity eigenvalues experience discrete rotations in the complex plane, described by the topological bulk indices $w = \frac{1}{2}\left[ {1 - {\rm sgn} ( \eps  ),1 - {\rm sgn} ( \mu)} \right]$, and the difference of these $\mathbb{Z}_2$ indices between the two media provide the number of surface electromagnetic modes (one/two modes when one/two indices change), Fig.~\ref{Fig1}(a).

Notably, the equations for sound waves in fluids or gases also have a form similar to Maxwell equations and also involve two medium parameters: the density $\rho$ and compressibility $\beta$. Many electromagnetic and acoustic quantities (e.g., the energy density, energy flux density, etc.) have similar forms described by the substitution of the Maxwell (electric, magnetic) and acoustic (velocity, pressure) fields: $\left( {{\bf{E}},{\bf{H}}} \right) \leftrightarrow \left( {{\bf{v}},P} \right)$, as well as the medium parameters: $\left( {\varepsilon ,\mu } \right) \leftrightarrow \left( {\rho ,\beta } \right)$ \cite{Bliokh2019II,Bliokh2019III}. However, surprisingly, surface acoustic waves exist only at interfaces where $\rho$ changes its sign (but no additional mode appears at interfaces where $\beta$ changes its sign) \cite{Ambati2007,Christensen2007,Park2011,Bliokh2019II}, Fig.~\ref{Fig1}(b). This breaks the symmetry between acoustic and Maxwell equations.

In this paper, we explain this enigmatic property by revealing the fundamental topological origin of surface acoustic waves. Akin to the Maxwell case \cite{Bliokh2019}, this requires representing the equations of acoustics in the fundamental field-theory form. However, since phonons are spinless particles, this corresponds to the {\it Klein-Gordon (KG)} rather than the Weyl equation. \CKB{In contrast to previous Klein-Gordon approaches to acoustics \cite{Forbes2003,Forbes2004,Deymier2016,Deymier2018}, here we focus on the ``relativistic'' {\it four-momentum} operator in the problem rather than on the wave equation itself. Remarkably, we find that this operator is generally {\it non-Hermitian} (even in idealized lossless media with real-valued parameters) and it provides a single $\mathbb{Z}_2$ topological bulk index determined by ${\rm sgn} (\rho)$. According to this, surface waves are also generally ``non-Hermitian'', i.e., can have either real or imaginary frequency and propagation constant.}
A comparison between the electromagnetic Weyl and acoustic KG formalisms is provided in Supplemental Material \cite{SM}.

{\it Klein-Gordon form of acoustic wave equations.---}
We start with the acoustic wave equations for the real-valued pressure, $P( {{\bf{r}},t})$, and velocity, ${\bf v}( {{\bf{r}},t})$, fields \cite{LLFluid}:
\begin{equation}
\label{Eq.1}
\beta\, \frac{{\partial P}}{{\partial t}} =  - \nabla  \cdot {\bf{v}}~,\qquad
\rho\, \frac{{\partial {\bf{v}}}}{{\partial t}} =  - \nabla P~.
\end{equation}
These equations obey an acoustic analogue of the electromagnetic Poynting theorem: 
${\partial _t}W + \nabla  \cdot {\bm \Pi}  = 0$, where 
\begin{equation}
\label{Eq.2}
W = \frac{1}{2}\left( {\beta {P^2} + \rho {{\bf{v}}^2}} \right) \quad {\rm and} \quad
{\bm \Pi}  = P\, {\bf{v}}
\end{equation}
are the acoustic energy density and energy flux density, respectively. 

On the one hand, Eqs.~(1) describe {\it vector} waves, characterized by one scalar (pressure) and one vector (velocity) fields. These scalar and vector degrees of freedom are equally important, as can be seen from their symmetric contributions to the conserved quantities (2). In quantum-like terminology, one can say that acoustic waves are described by the four-component ``wavefunction'' ${\Psi ^\mu } = ({P,{\bf{v}}})$. On the other hand, the acoustic waves are longitudinal waves, which correspond to {\it spinless} quantum particles: phonons. It is known from field theory that spinless particles are described by a {\it single scalar field} and the {\it KG equation}.

The vector acoustic theory can indeed be reduced to the KG description using a single scalar field $\psi ({{\bf{r}},t})$. We employ the standard KG field Lagrangian ${\cal L} = \frac{1}{2}\left[ {{c^{ - 2}}{{\left( {{\partial _t}\psi } \right)}^2} - {{\left( {\nabla \psi } \right)}^2}} \right]$, where ${c^2} = {( {\rho \beta } )^{ - 1}}$ is the squared speed of sound. The equation of motion, energy density, and energy flux density for this Lagrangian read \cite{BLP}: 
\begin{gather}
\label{Eq.3}
{c^{ - 2}}\partial _t^2\psi  - {\nabla ^2}\psi  = 0~, \\
\label{Eq.4}
W = \frac{1}{2}\left[ {{c^{ - 2}}{{\left( {{\partial _t}\psi } \right)}^2} + {{\left( {\nabla \psi } \right)}^2}} \right], \quad
{\bm \Pi}  =  - {\partial _t}\psi \,\nabla \psi~.
\end{gather}
Comparing the vector acoustic equations (1) and (2) with the scalar KG equations (3) and (4), we find that these are equivalent, when the vector and scalar representations are related by $P =  - \sqrt \rho  \,{\partial _t}\psi$ and ${\bf{v}} = \left( {1/\sqrt \rho  } \right)\nabla \psi$. Thus, the KG wavefunction is similar to the scalar velocity potential $\varphi$, ${\bf{v}} = \nabla \varphi$, widely used in acoustics \cite{LLFluid}, but the pre-factors involving $\sqrt{\rho}$ will play a crucial role in our theory below. 

To provide an elegant representation of the KG formalism, we introduce ``four-vectors'' ${a^\mu } = \left( {{a^{(P)}},\,{{\bf{a}}^{({\bf{v}})}}} \right)$, ${a_\mu } = \left( {{a^{(P)}},\, - {{\bf{a}}^{({\bf{v}})}}} \right)$, where the time-like and space-like components are related to the pressure and velocity degrees of freedom, and the metric with signature $( { + , - , - , - })$ is used. Of course, the acoustic equations (1) are {\it not} relativistic and Lorentz-covariant, but using the relativistic formalism, natural for the KG equation, facilitates the consideration. To describe scalar and three-vector properties, we introduce the following bilinear operations for four-vectors:
\begin{gather}
\label{Eq.5}
 {a^\mu}\bullet{b_\mu } \equiv \beta \,{a^{(P)}}{b^{(P)}} - \rho \,{{\bf{a}}^{({\bf{v}})}} \cdot {{\bf{b}}^{({\bf{v}})}}, \nonumber\\
 {a^\mu } \otimes {b_\mu } \equiv {a^{(P)}}{{\bf{b}}^{({\bf{v}})}} - {{\bf{a}}^{({\bf{v}})}}{b^{(P)}}~.
\end{gather}
The first operation (5) is a scalar product modified by the scaling coefficients $\beta$ and $\rho$, to provide the correct dimensionality, while the second operation is a ``cross-product'', which produces a three-vector (note that ${a^\mu } \otimes {a_\mu } = 0$).

Using these notations, the KG equation and the relation between the KG wavefunction and physical fields can be written as:
\begin{gather}
\label{Eq.6}
 \left( {{{\hat p}^\mu}\bullet{{\hat p}_\mu }} \right)\psi  = \left( { - {c^{ - 2}}\partial _t^2 + {\nabla ^2}} \right)\psi  = 0~, \\
 {\Psi ^\mu }\! = i{\hat p^\mu } \psi,~~
{\hat p^\mu } = \left( {i\sqrt \rho\,  {\partial _t},\, \frac{{ - i\nabla }}{{\sqrt \rho  }}} \right),
~~
{\Psi ^\mu }\! \equiv \left( {P,\, {\bf{v}}} \right).
\end{gather}
Here, the operator $\hat{p}^\mu$ should be associated with the {\it four-momentum} in the problem. Equations of motion (1) also take a very laconic form:
\begin{equation}
\label{Eq.8}
{\hat p^\mu } \bullet {\Psi _\mu } = 0~,  \qquad
{\hat p^\mu } \otimes {\Psi _\mu } = {\bf{0}}~.
\end{equation}
Here the first equation is obtained from Eqs.~(6) and (7), while the second equation is a consequence of Eq.~(7) and $\hat{p}^\mu \otimes \hat{p}_\mu  = 0$. Thus, the acoustic KG equation (6) uses only the first equation of motion (1), while the second equation of motion (yielding $\nabla  \times {\bf{v}} = {\bf 0}$) follows from the definitions (${\bf{v}} \propto \nabla \psi$). This is similar to the Weyl-like form of Maxwell equations, where equations $\nabla  \cdot {\bf{E}} = 0$ and $\nabla \cdot {\bf{H}} = 0$ are not used \cite{Bliokh2019} (see Supplemental Material \cite{SM}). Finally, the acoustic energy density and energy flux, Eqs.~(2) and (4), also take simple forms: 
\begin{equation}
\label{Eq.9}
W  = \frac{1}{2}{\Psi ^\mu } \bullet {\Psi ^\mu },  \qquad
{\bm \Pi}  = \frac{1}{2}{\Psi ^\mu } \otimes {\Psi ^\mu }.
\end{equation}

Equations (3)--(9) provide the KG representation of acoustic equations (1) and (2), both in terms of the scalar KG wavefunction $\psi$ and vector wavefunction $\Psi^mu$ involving real physical fields (cf. the electromagnetic Weyl representration in Supplemental Material \cite{SM}). Both of these are important, because the real physical fields $( {P,{\bf{v}}})$ determine the boundary conditions (i.e., the corresponding components are continuous at interfaces), while the KG wavefunction $\psi$ provides the fundamental field-theory representation of the problem. One can see that Eqs.~(3)--(9) are {\it not} relativistic-invariant, because the only relativistic-invariant four-vector equations in the Klein-Gordon problem are \cite{BLP}: ${\hat p^\mu }\psi  = m\, {\Psi ^\mu }$, ${\hat p^\mu } {\Psi _\mu } = m\,\psi$, $\left( {{{\hat p}^\mu }\, {{\hat p}_\mu } - {m^2}} \right)\psi  = 0$, where $m$ is the mass, which obviously differs from Eqs.~(6)--(8). 


{\it Topological non-Hermitian origin of surface modes.---}
The most remarkable feature of the above formalism is contained in Eq.~(7). Namely, while all other equations are fairly symmetric with respect to the $\rho$ and $\beta$ parameters, the {\it four-momentum operator $\hat{p}^\mu$ involves only the density $\rho$ but not the compressibility $\beta$} (cf. the electromagnetic four-momentum or helicity involving both $\eps$ and $\mu$ in Supplemental Material \cite{SM}). This operator is the key operator in the KG formalism, and it is this operator (rather than the ``Hamiltonian'' $\hat{p}^\mu \bullet \hat{p}_\mu$) which should be considered for the topological classification of acoustic media. 

Remarkably, the four-momentum operator is generally {\it non-Hermitian} with respect to the standard inner product, because in negative density media, $\rho <0$, it has purely {\it imaginary} eigenvalues. In complete analogy with the non-Hermitian helicity operator for photons \cite{Bliokh2019,Alpeggiani2018}, the $\hat{p}^\mu$ operator is singular at $\rho=0$, which separates topologically-different phases with real ($\rho >0$) and imaginary ($\rho <0$) eigenvalues. 
\CKB{It should be emphasized that the ``anomaly'' of the four-momentum in negative-density media is not due to the imaginary wavevector ${\bf k}$ or frequency $\omega$ (which also occur in negative-$\beta$ media), but due to imaginary proportionality factors between the four-momentum eigenvalues $p^\mu$ and the four-wavevector $k^\mu = (\omega, {\bf k})$. The same topological difference between the $\rho>0$ and $\rho<0$ zones, separated by the $\rho=0$ singularity, appears in the connection between the Klein-Gordon wavefunction $\psi$ and real physical fields $(P,{\vec v}) = i\hat{p}^\mu \psi$.}

Then, entirely similar to \cite{Bliokh2019} and a number of recent results on the topological properties of non-Hermitian systems \cite{Leykam2017,Kunst2018,Yao2018,Gong2018}, we note that the topologically-different cases of purely-real and purely-imaginary four-momentum eigenvalues can be labeled by the $\mathbb{Z}_2$ {\it topological bulk invariant} for acoustic media:
\begin{equation}
\label{Eq.10}
w( \rho ) = \frac{1}{2}\left[ {1 - {{\rm sgn}} ( \rho )} \right] \in \{0,1\}.
\end{equation}
Furthermore, according to the {\it bulk-boundary correspondence}, an interface between media ``1'' and ``2'' supports $N$ surface modes, where
\begin{equation}
\label{Eq.11}
N = \left| {w ( {{\rho _2}} ) - w ( {{\rho _1}} )} \right| = \left| {w ( {{\rho _r}} )} \right| \in \{0,1\},
\end{equation}
where ${\rho _r} = {\rho _2}/{\rho _1}$ is the relative density of the media. This means that {\it there is a single surface acoustic mode at interfaces with ${{\rm sgn}} ( {{\rho _r}} ) =  - 1$} (topologically-different media) and {\it no surface modes at interfaces with ${{\rm sgn}} ( {{\rho _r}} ) = 1$} (topologically-equivalent media), as shown in Fig.~\ref{Fig1}(b). These findings are in agreement with the topological properties of various non-Hermitian systems \cite{Leykam2017,Kunst2018,Yao2018,Gong2018} where topological transitions occur at exceptional points where the spectrum of the non-Hermitian operator changes from real to imaginary. The only principal difference here is that most of the previous studies considered non-Hermitian {\it Hamiltonian} operators \CKB{(i.e. systems with physical losses or gain)}, while here, similar to \cite{Bliokh2019} (see also Supplemental Material \cite{SM}), we use another key operator, the {\it four-momentum} of phonons. \CKB{Akin to the helicity of photons, this operator can be mathematically non-Hermitian even in idealized lossless systems.}

An important feature of the non-Hermitian topological approach is that the surface modes are also generally ``{\it non-Hermitian}''. This means that these can have complex (actually, either real or imaginary) frequencies and/or propagation constants. We will refer to surface modes with real and imaginary frequency/propagation characteristics as propagating and evanescent surface waves, respectively. For a planar interface between two media, the parameters of the surface acoustic mode are determined by the following equations \cite{Ambati2007,Christensen2007,Bliokh2019II}:
\begin{gather}
\label{Eq.12}
 \frac{{{\kappa _1}}}{{{\rho _1}}} + \frac{{{\kappa _2}}}{{{\rho _2}}}  = 0~, \\
\label{Eq.13}
 k_{\rm surf}^2 = \kappa _1^2\frac{{{\rho _r}\left( {{\beta _r} - {\rho _r}} \right)}}{{{\rho _r}{\beta _r} - 1}}~, ~~
\omega _{\rm surf}^2 = c_1^2\kappa _1^2\frac{{\left( {1 - \rho _r^2} \right)}}{{{\rho _r}{\beta _r} - 1}}~.
\end{gather}
where $\kappa_{1,2}>0$ are the spatial-decay (away from the interface) constants in the two media, $\omega _{\rm surf}$ and $k_{\rm surf}$ are the frequency and propagation constant of the surface wave, ${c_1} = 1/\sqrt {{\rho _1}{\beta _1}}$ is the speed of sound in the first medium, and ${\beta _r} = {\beta _2}/{\beta _1}$ is the relative compressibility. Equation (12) requires ${\rho _r} < 0$, which is exactly the topological condition (10) and (11). This condition determines the {\it domain of existence} of the surface mode, Fig.~\ref{Fig1}(b), which is {\it robust (topologically protected) and independent of the shape of the interface}. Note that Eq.~(12) actually follows from the second equation (1) (which determines the longitudinal character of sound waves, $\nabla  \times {\bf{v}} = {\bf 0}$) and the continuity of the pressure and normal velocity component. At the same time, Eqs.~(13) determine {\it specific parameters} of the surface mode, which are {\it not robust} and can vary with the shape of the interface.

\begin{figure}
\includegraphics[width=\columnwidth]{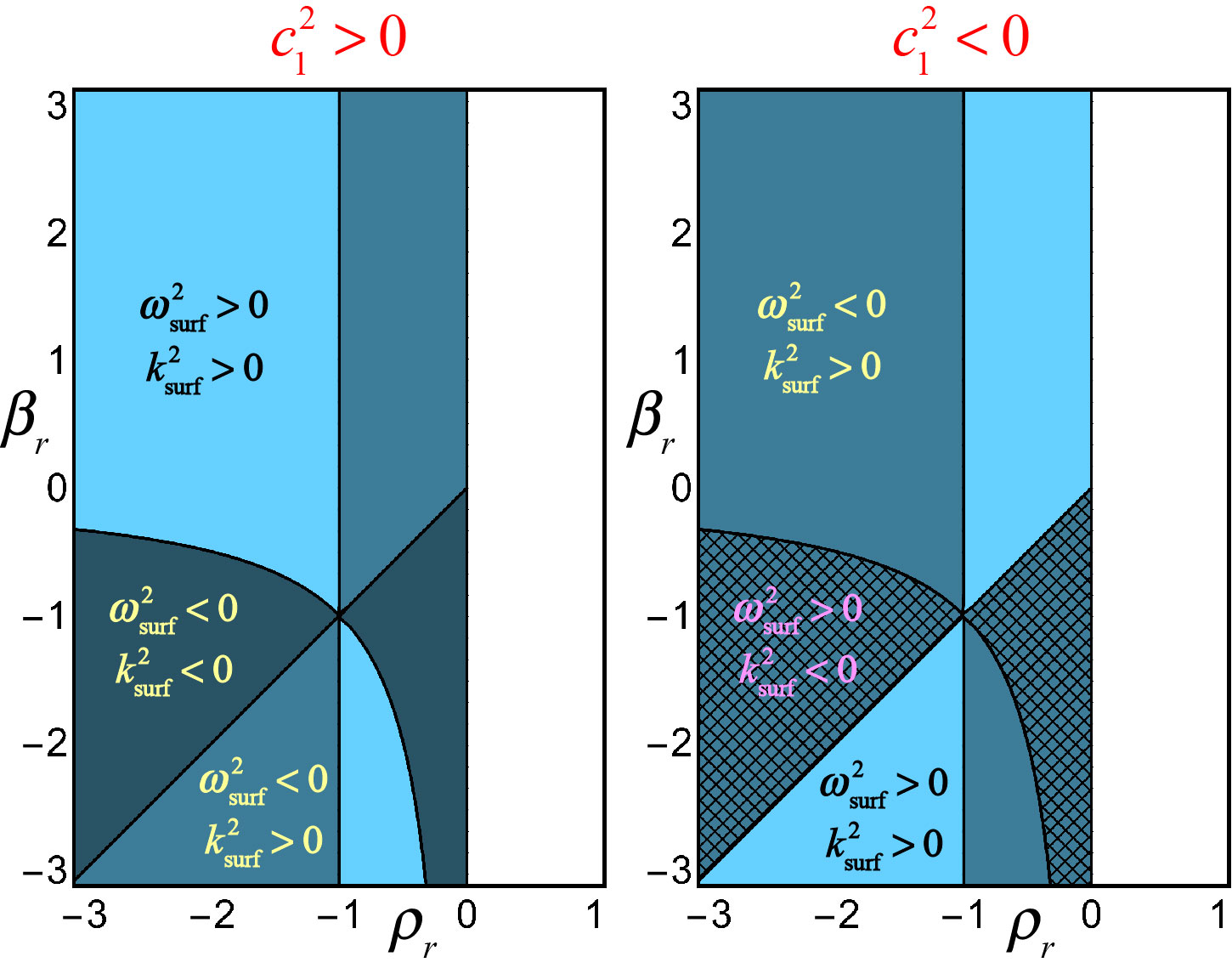}
\caption{Due to the non-Hermitian origin of surface acoustic modes, the global domain of their existence, Fig.~\ref{Fig1}(b), is split into zones with real and imaginary frequencies and propagations constants. These zones differ for the cases of a transparent ($c_1^2 >0$) and non-transparent ($c_1^2 <0$) first medium. The light-blue zones correspond to the usual case of propagating surface modes with real frequency and propagation constant \cite{Bliokh2019II}. The separation of the propagating and evanescent surface-wave zones is not robust (topologically protected) and can vary with properties of the interface: its shape, dispersion of the media, etc.  
\label{Fig2}}
\end{figure}

Figure~\ref{Fig2} shows the zones of propagating ($\omega _{\rm surf}^2 > 0$, $k _{\rm surf}^2 > 0$) and evanescent ($\omega _{\rm surf}^2 < 0$ or/and $k _{\rm surf}^2 < 0$) surface acoustic modes, which follow from Eqs.~(13). Usually, only propagating surface modes are considered and observed experimentally. Moreover, evanescent modes with imaginary frequencies ($\omega _{\rm surf}^2 < 0$) cannot exist in the physical lossless systems considered here \cite{Silveirinha}. Indeed, in lossless media, the frequency spectrum must be symmetric with respect to the ${\rm Re}\omega$ axis, but exponentially-growing solutions with ${\rm Im}\omega >0$ cannot exist in systems without gain. Nonetheless, these solutions are formally present in the acoustic equations for monochromatic fields ${\bf{v}}( {{\bf{r}},t} ) \to {\rm Re}\! \left[ {\bf{v}}( {\bf{r}}){e^{ - i\omega t}} \right]$ and $P( {{\bf{r}},t} ) \to {\rm Re}\! \left[ P( {\bf{r}}){e^{ - i\omega t}} \right]$, where ${\bf{v}}( {\bf{r}})$ and $P( {\bf{r}})$ are now complex field amplitudes. Notably, there are also evanescent surface modes with real frequencies and imaginary propagation constants ($\omega _{\rm surf}^2 > 0$, $k _{\rm surf}^2 < 0$), which can appear in reality. Such modes can occur at interfaces between two non-transparent media with opposite-sign densities and compressibilities: $c_{1,2}^2<0$, $\rho_r <0$, $\beta_r <0$. 

It should be emphasized that here we deal with an idealized situation of {\it lossless} and {\it non-dispersive} media. In reality, negative parameters, necessary for the appearance of surface modes, can be achieved only in acoustic metamaterials, i.e., {\it dispersive} media. This also leads to inevitable presence of losses. Therefore, real physical systems with surface modes should be described by a complex frequency-dependent density $\rho(\omega)$ and compressibility $\beta(\omega)$. Nonetheless, in many cases, the diagrams obtained for lossless non-dispersive media are sufficient to obtain the main properties of surface modes, such as surface plasmon-polaritons \cite{Zayats2005,Maier2007,Shadrivov2004,Kats2007}. Indeed, Eqs.~(\ref{Eq.12}) and (\ref{Eq.13}) remain valid with the substitution $\{\rho,\beta\} \to \{{\rm Re}\rho(\omega),{\rm Re}\beta(\omega)\}$, provided that the imaginary parts of these parameters are negligible. In this case, the dispersion can only modify the parameters of the non-Hermitian surface modes shown in Fig.~\ref{Fig2}, but it does not affect the fundamental topological origin of these modes, Eqs.~(\ref{Eq.10})--(\ref{Eq.12}) and Fig.~\ref{Fig1}(b).  
\CKB{This is because the dispersion only modifies the {\it inner product} of the wavefunction in the problem, but does not affect the {\it operators}  which are reponsible for the topological classification of media \cite{Alpeggiani2018, Silveirinha2015, Silveirinha2017, Bliokh2017PRL, Bliokh2019}.}


{\it Discussion.---}
We have shown that equations for acoustic waves in fluids or gases allow the relativistic-like Klein-Gordon representation. This representation breaks the symmetry between the pressure and velocity degrees of freedom and is characterized by the nontrivial ``four-momentum'' operator. Importantly, in contrast to other fundamental quantities, this operator depends only on one parameter (density $\rho$ but not compressibility $\beta$), and it is generally non-Hermitian \CKB{(even in idealized lossless media with real parameters)}. The $\rho=0$ point splits the parameter space into two topologically-different phases with real ($\rho>0$) and imaginary ($\rho<0$) four-momentum eigenvalues. This allows one to introduce the $\mathbb{Z}_2$ bulk topological index (10), the bulk-boundary correspondence (11), and reveals the topological origin of acoustic surface waves (analogues of surface plasmon-polaritons in electromagnetism), Fig.~\ref{Fig1}(b). Remarkably, the non-Hermitian nature of the four-momentum operator leads to non-Hermitian surface modes, which can have either real or imaginary frequencies and/or propagation constants. This further splits the simple phase diagram Fig.~\ref{Fig1}(b) into zones of propagating and evanescent surface modes, Fig.~\ref{Fig2}. However, this splitting is not topologically protected and can vary with the shape of interface and other perturbations.

This work explains the fundamental origin of acoustic surface waves, which is present already in the simplest case of nondispersive and lossless media. 
In practice, negative medium parameters can be achieved only in dispersive, and therefore lossy, metamaterials. In the case of small losses, the dispersion can modify details of the non-Hermitian diagrams in Fig.~\ref{Fig2}, but it cannot affect the fundamental topological properties, Eqs.~(\ref{Eq.10})--(\ref{Eq.12}) and Fig.~\ref{Fig1}(b). 
Still, an accurate generalization of the topological consideration to dispersive lossy media is an important task for future studies, both in acoustics and electromagnetism.
The comparison of acoustic and electromagnetic surface waves provides insights for both theories, highlighting their similarities and distinctions. Both electromagnetic and acoustic equations can be presented in the form of relativistic wave equations: the Weyl and Klein-Gordon ones (see Supplemental Material \cite{SM}). Furthermore, the main operators of these equations, helicity and four-momenta, are non-Hermitian and their topological phases determine the phase diagrams of surface modes, Fig.~\ref{Fig1}. There is one more classical wave theory: elasticity. It combines features of acoustics and electromagnetism, because there are both longitudinal (sound) and transverse (shear, photon-like) modes. Constructing the topological theory for elastic waves is another important and challenging problem.


\vspace*{0.2cm}

\begin{acknowledgments}
We are grateful to D. Leykam, M. Lein, M. Silveirinha, and J. Dressel for fuitful discussions. This work was partially supported by MURI Center for Dynamic Magneto-Optics via the Air Force Office of Scientific Research (AFOSR) (FA9550-14-1-0040), Army Research Office (ARO) (Grant No. Grant No. W911NF-18-1-0358), Asian Office of Aerospace Research and Development (AOARD) (Grant No. FA2386-18-1-4045), Japan Science and Technology Agency (JST) (Q-LEAP program, and CREST Grant No. JPMJCR1676), Japan Society for the Promotion of Science (JSPS) (JSPS-RFBR Grant No. 17-52-50023, and JSPS-FWO Grant No. VS.059.18N), RIKEN-AIST Challenge Research Fund, the John Templeton Foundation, and the Australian Research Council.
\end{acknowledgments}

\bibliography{bibA}

\newpage






\begin{figure}
 \centering 
 \includegraphics[scale=0.8]{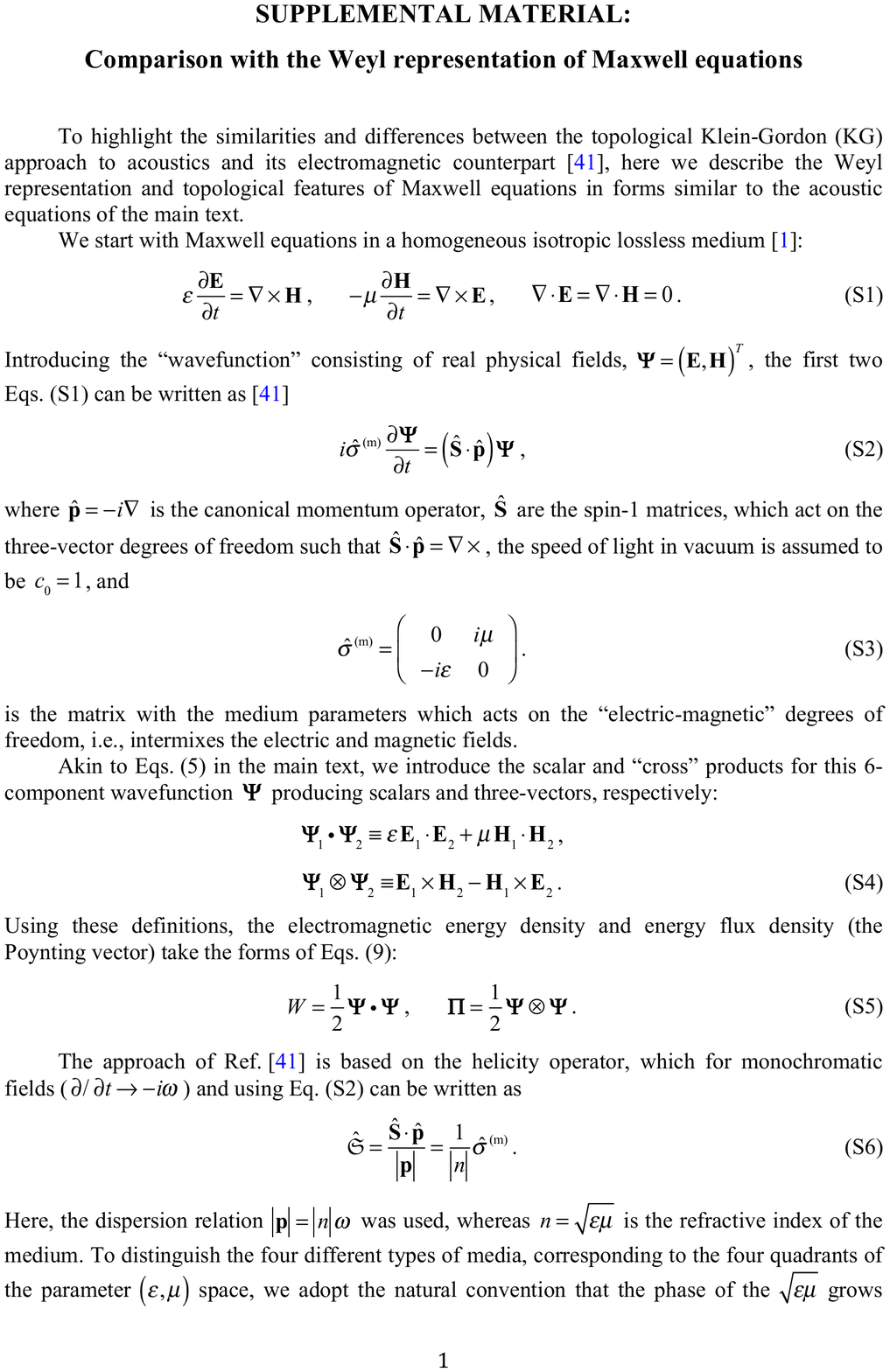}
\end{figure}

\begin{figure}
 \centering 
 \includegraphics[scale=0.8]{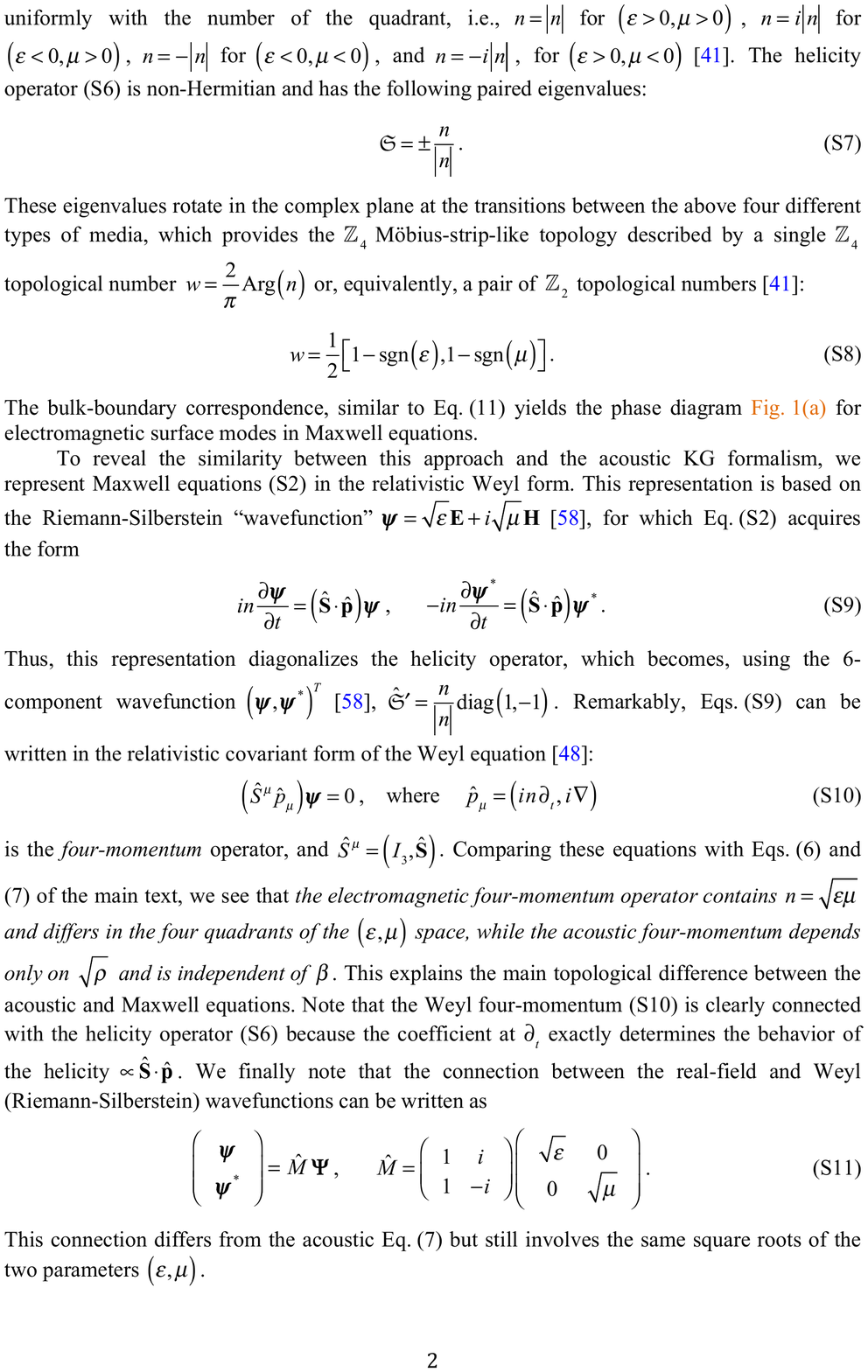}
\end{figure}

\begin{figure}
 \centering 
 \includegraphics[scale=0.8]{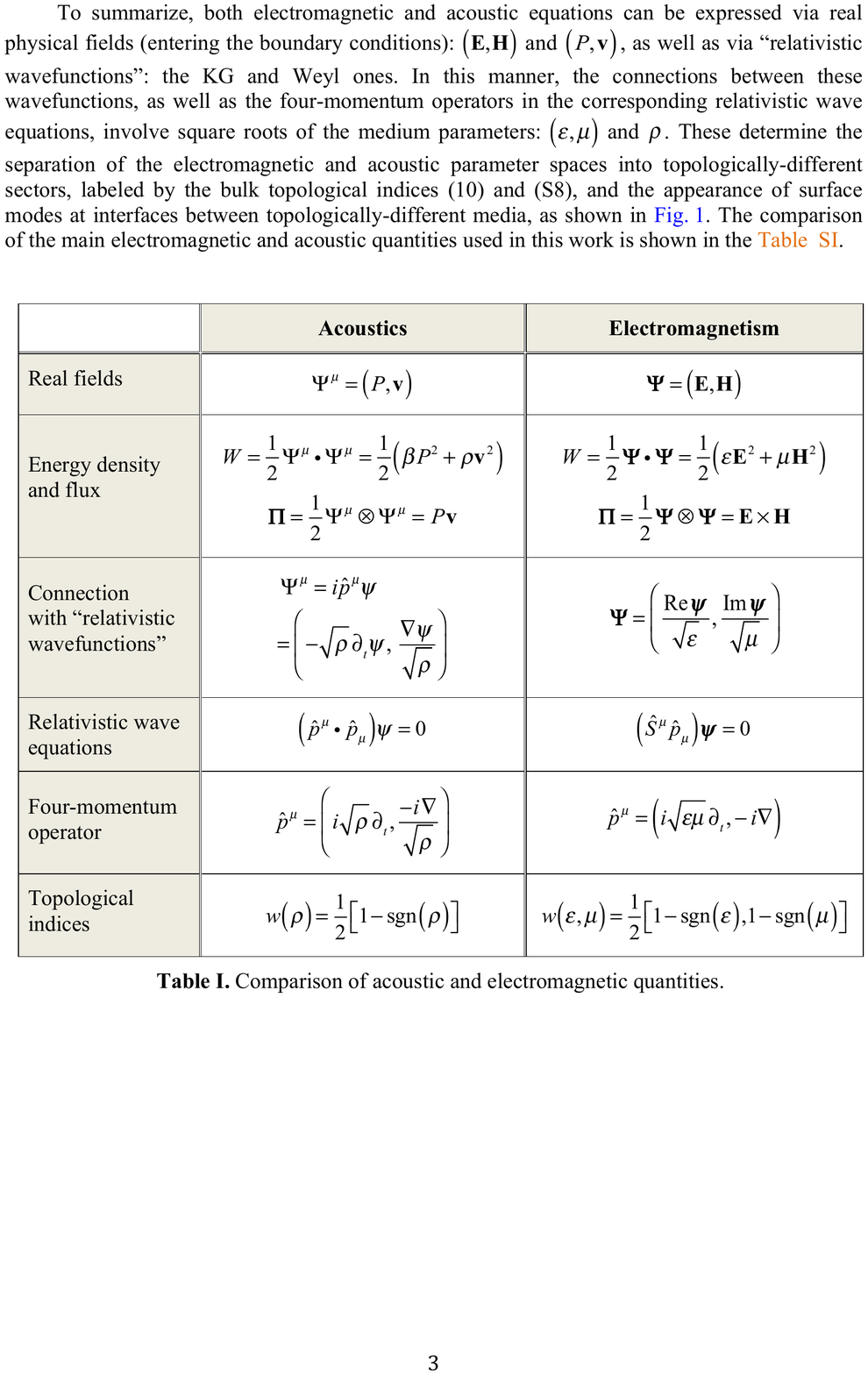}
\end{figure}

\end{document}